\documentclass[english,journal=jpcafh,manuscript=article,email=true,etalmode=truncate,maxauthors=0,doi=true]{achemso}

\SectionNumbersOn

\usepackage{graphicx}
\graphicspath{{figures/}}
\usepackage{physics}
\usepackage{amsmath}
\usepackage{multirow}
\usepackage{xcolor}
\usepackage{caption}
\usepackage{subcaption}
\usepackage{xcolor}
\usepackage{bm}
\usepackage{eufrak} % for \mathfrak
\usepackage{amssymb} % for \mathbb
\usepackage[title,titletoc]{appendix}
\usepackage{diagbox}
\usepackage{multicol}

\usepackage[numbers,sort&compress,super]{natbib}
\usepackage{natmove}
\usepackage{hyperref} % hyperref must be placed before cleveref
\hypersetup{colorlinks=true, citecolor=blue, urlcolor=blue, linkcolor=blue}
\usepackage{cleveref}
  	\crefname{figure}{Figure}{Figures}
  	\crefname{table}{Table}{Tables}
  	\crefname{equation}{Eq.}{Eqs.}
  	\crefname{section}{Section}{Sections}
  	\crefname{subsection}{Section}{Sections}
  	\crefname{subsubsection}{Section}{Sections}
  	\crefname{algorithm}{Algorithm}{Algorithms}

\newcommand{\doi}[1]{\href{http://dx.doi.org/#1}{\nolinkurl{#1}}}

\usepackage[section,frozencache]{minted}      % change 'finalizecache' to 'frozencache' before submitting to arXiv

\newcommand{\code}[1]{\texttt{#1}}

\usepackage{todonotes}

\title{High-performance evaluation of high angular momentum 4-center Gaussian integrals on modern accelerated processors} %Title of paper

\author{Andrey Asadchev}

\author{Edward F. Valeev}
\affiliation{Department of Chemistry, Virginia Tech, Blacksburg, VA 24061}
\email{efv@vt.edu}

\begin{document}

\date{\today}

\begin{abstract}
We present a high-performance evaluation method for 4-center 2-particle integrals over Gaussian atomic orbitals with high angular momenta ($l\geq4$) and arbitrary contraction degrees on graphical processing units (GPUs) and other accelerators. The implementation uses the matrix form of McMurchie-Davidson recurrences. Evaluation of the 4-center integrals over four $l=6$ ($i$) Gaussian AOs in the double precision (FP64) on an NVIDIA V100 GPU outperforms the reference implementation of the Obara-Saika recurrences (\code{Libint}) running on a single Intel Xeon core by more than a factor of 1000, easily exceeding the 73:1 ratio of the respective hardware peak FLOP rates while reaching almost 50\% of the V100 peak. The approach can be extended to support AOs with even higher angular momenta; for lower angular momenta ($l\leq3$) additional improvements will be reported elsewhere. The implementation is part of an open-source \code{LibintX} library feely available at \url{github.com:ValeevGroup/LibintX}.
\end{abstract}

\maketitle %\maketitle must follow title, authors, abstract

\section{Introduction}\label{sec:intro}
Efficient evaluation of 4-center 2-particle integrals over Gaussian AOs with high angular momenta is a major challenge for modern high-performance computing (HPC) platforms. Traditional evaluation schemes, such as recurrences, involve custom kernels with irregular compute and data access patterns that do not map well onto the modern hardware that has evolved to execute optimally data-parallel code with regular instruction and data access structure (such as the dense matrix multiplication). An even more serious issue is the high memory footprint of such kernels that exceeds the size of the lowest levels of memory hierarchy (registers and {\em scratchpad} memory\footnote{Scratchpad memory is known to as shared memory in the CUDA programming model and local data store in the HIP programming model}) even for relatively low angular momenta, thereby reducing the performance. Although detailed performance can be difficult to extract from the numerous publications dedicated to Gaussian AO integral evaluation on GPUs,\cite{VRG:yasuda:2007:JCC,VRG:ufimtsev:2009:JCTC,VRG:miao:2013:JCTC,VRG:yasuda:2014:IJQC,VRG:miao:2015:JCTC,VRG:rak:2015:CPL,VRG:kalinowski:2017:JCTC,VRG:song:2016:JCTC,VRG:kussmann:2017:JCTC,VRG:tornai:2019:JCTC,VRG:barca:2020:JCTC,VRG:barca:2020:2SICHPCNSAS,VRG:barca:2021:JCTC,VRG:barca:2021:PICHPCNSA,VRG:johnson:2022:JCTC,VRG:galvezvallejo:2022:MP,VRG:asadchev:2023:JCTC} the performance of the Head-Gordon-Pople\cite{VRG:obara:1986:JCP,VRG:head-gordon:1988:JCP} refinement of the Obara-Saika\cite{VRG:obara:1986:JCP} recurrence scheme implemented by Barca et al.\cite{VRG:barca:2021:JCTC} is in our experience representative: whereas the performance for 4-center integrals for low total angular momenta (up to $(pp|pp)$, with $s,p,d,f,g,h,i,k\dots$ denoting Gaussian AOs with angular momenta $l=0,1,2,3,4,5,6,7\dots$, respectively) was found to reach a substantial (20-50\%) fraction of the peak FP64 FLOP rate, the performance for higher angular momenta dropped rapidly, to 2\% of the peak rate for the $(dd|dd)$ integrals. Another, albeit, a less direct, datapoint comes from a study by Johnson et al.\cite{VRG:johnson:2022:JCTC} who observed significant loss of efficiency of the GPU code for the Coulomb matrix evaluation (using McMurchie-Davidson recurrence-based formalism\cite{VRG:mcmurchie:1978:JCP}) vs the CPU counterpart as the basis set is enlarged to include higher angular momenta. Specifically, the GPU vs. single CPU core speedup in the the cc-pV\{D,T,Z\}Z basis set series drops from $\sim50$ to $\sim10$ for the largest test cited in Figure 9 of Ref. \citenum{VRG:johnson:2022:JCTC}.

Recently we reconsidered the design of Gaussian AO integral algorithms in order to optimize their memory footprint.\cite{VRG:asadchev:2023:JCTC} For the specific case of 3-center Gaussian AO integrals we argued that even for high angular momenta the Obara-Saika recurrence-based schemes\cite{VRG:obara:1986:JCP,VRG:ahlrichs:2006:PCCPP} would outperform the Rys quadrature\cite{VRG:dupuis:1976:JCP,VRG:rys:1983:JCC} commonly thought to lead to optimally-compact memory footprints; however even with several algorithmic and programming innovations the performance was reasonable for integrals up to $(ff|f)$ but dropped for higher angular momenta to only a few percent of the hardware peak. Straightforward application of these ideas to 4-center integrals would surely lead to similar performance that would be adequate for relatively low angular momenta. Clearly, a new line of attack was needed.

A possible workaround for the challenge of high-$l$ integrals is the use of real-space factorization of 2-electron integrals, as illustrated recently by us and collaborators\cite{VRG:williams-young:2023:JCP} via the use of real-space quadrature (``pseudospectral''\cite{VRG:friesner:1985:CPL}, also known as chain-of-spheres\cite{VRG:neese:2009:CP} or seminumerical\cite{VRG:laqua:2018:JCTC}) approximation to the exact exchange which trades the problem of computing 4-center 2-electron integrals for evaluation of cheaper but much more numerous 2-center 1-electron Gaussian AO integrals. Nevertheless, these and other numerical approximations that avoid the 4-center integrals cannot entirely eliminate the need to evaluate 4-center 2-electron integrals, thus their efficient evaluation, especially for high angular momenta, remains a critical challenge on modern HPC platforms.

A conventional wisdom then calls for the use of the Rys quadrature\cite{VRG:dupuis:1976:JCP,VRG:rys:1983:JCC} since its memory footprint for the high angular momentum integrals is smaller than that of the recurrence-based schemes. The peak memory footprint of the Rys-based evaluation are the 4-index (1-dimensional) integrals that for a 4-center integral over Gaussians with angular momentum $l$ require $3 \times (l+1)^4 \times N_\mathrm{roots}$ words, where $N_\mathrm{roots} = 2l+1$ is the number of the quadrature roots.  While these intermediates would fit within the 96 kB maximum size of a single shared memory block of the NVIDIA V100 GPU for the $(ff|ff)$ shell set, the $(gg|gg)$ case (135 kB) would not. The latter would fit into the larger maximum shared memory on more recent A100 and H100 devices from NVIDIA (164 kB and 228 kB, respectively). However, there is only enough memory for one thread block, rather than two or more needed to be able to hide latencies and stalls. Dealing with contracted Gaussians will increase the required memory requirements further. Lastly, the Rys quadrature is more difficult to generalize to non-Coulomb operators than the alternatives.

Thus the question remains: what is a better way to evaluate 2-electron integrals over high-$l$ Gaussian AOs on GPUs? Our recent work\cite{VRG:williams-young:2023:JCP} on the density-fitting-accelerated J-matrix engine\cite{VRG:shao:2000:CPL} implementation for GPUs based on McMurchie-Davidson (MD) recurrences\cite{VRG:mcmurchie:1978:JCP}
hinted that the MD recurrences recast in matrix form might be the way to go for some classes of integrals; that development also provided many fundamental elements that we reused in this work. We were not alone in thinking that: the \code{SHARK} integral engine developed by Frank Neese\cite{VRG:neese:2022:JCC} and recently incorporated into the public release of \code{ORCA} program\cite{VRG:neese:2020:JCP} illustrated how efficient the MD scheme can be when expressed as a matrix multiplication ({\em matmul}) on conventional CPUs, at least when used for integrals over high angular momenta (\code{SHARK} supplements the MD approach with traditional Obara-Saika-based kernels implemented in the \code{Libint} library).

The desire to recast integral evaluation in terms of matrix/tensor kernels is not recent; for example, the COLD PRISM of Gill et al.\cite{VRG:adams:1997:JCP} was formulated  and implemented explicitly in terms of BLAS-2 (matrix $\times$ vector) and BLAS-3 (matrix $\times$ matrix) operations. In a sense, the density fitting\cite{VRG:whitten:1973:JCP,VRG:harris:1966:TCA,VRG:baerends:1973:CP} and other related factorizations\cite{VRG:beebe:1977:IJQC, VRG:hohenstein:2012:JCP, VRG:pierce:2021:JCTC,VRG:friesner:1985:CPL} of the many-electron integrals can also be viewed as leveraging the matrix multiplication to work around the challenge of computing 4-center integrals.
The importance of dense matrix multiplication as the key building block has only risen recently due to the emergence of deep neural networks as the building blocks for data analytics and machine learning. Not only is dense matrix multiplication particularly efficient on modern HPC platforms, but modern commodity processors are being {\em engineered} to execute matrix multiplication as cost-effectively as possible. Engineering for matrix multiplication ranges from including co-processors for matrix multiplication (e.g., the AMX unit on Apple's ARM chips) to including dedicated functional units (tensor cores on NVIDIA GPUs) to purpose-built silicon (Google TPUs and other AI accelerators). Leveraging these trends by co-designing our algorithms to take advantage of the matrix hardware makes much sense.

In this manuscript we discuss the implementation and performance of the matrix-based formulation of the MD scheme for 4-center 2-electron Gaussian AO integrals on NVIDIA GPUs. The extensive use of generalized matrix multiply (GEMM) kernel is the key implementation detail that distinguishes our approach from the pioneering implementation of the MD scheme by Mart\'{i}nez et al.\cite{VRG:ufimtsev:2009:JCTC,VRG:johnson:2022:JCTC} Although the approach has been implemented to support 3-center integrals also, our focus here is squarely on the former; the 3-center case, being more (not less) complex,
will be presented elsewhere. The approach discussed here is applicable to any
reasonable combination of angular momentum and total contraction order $K$, for example $(hh|hh)$ with
$K=100$ or $(ff|ff)$ with $K=400$.  Note however that this approach is targeted
at moderately contracted high angular momentum integrals.  For low angular momentum and high
contraction order we will soon follow up with an improved variant.

\section{Formalism}

% McMurchie-Davidson (MD) scheme for evaluation of 4-center 2-electron integrals over Gaussian AOs consists of the following steps: (a) evaluation of Boys function values (for Coulomb kernels) or related quantities\cite{VRG:ahlrichs:2006:PCCPP} (for non-Coulomb kernels), (b) their conversion into 1-index (``center'') Hermite integrals, (c) conversion of the 1-index into the 2-index Hermite Gaussian integrals, and (d) conversion of the bra and ket from the Hermite to the target Cartesian or solid harmonic Gaussians.

Let's start with a quick review of the standard McMurchie-Davidson scheme for the evaluation of 4-center 2-electron integrals over Gaussian AOs (``MD4''); the reader is referred to the original reference\cite{VRG:mcmurchie:1978:JCP} and other sources\cite{VRG:helgaker:2000:,VRG:samu:2018:} for additional details.

Our notation closely follows that of Obara and Saika.\cite{VRG:obara:1986:JCP}
An uncontracted primitive Cartesian Gaussian with exponent $\zeta_a \in\mathbb{R}^+$ and non-negative integer Cartesian ``quanta'' $\mathbf{a} \equiv \{a_x, a_y, a_z \}$ centered at $\mathbf{R}_a \equiv \{A_x, A_y, A_z\}$ will be denoted by
\begin{align}
    \phi_\mathbf{a} ({\bf r}) \equiv x_A^{a_x} y_A^{a_y} z_A^{a_z} \exp(-\zeta_a r_A^2),
\end{align}
with $\mathbf{r}_A \equiv \{x_A, y_A, z_A\}, x_A \equiv x - A_x$, etc.
$l_\mathbf{a} \equiv a_x + a_y + a_z \geq 0$ is colloquially referred to as the ``angular momentum'' of a Gaussian. Closely related to the Cartesian Gaussian is a Hermite Gaussian:
\begin{align}
    \Lambda_\mathbf{\tilde{a}} ({\bf r}) \equiv \left(\frac{\partial}{\partial x_A}\right)^{\tilde{a}_x} \left(\frac{\partial}{\partial y_A}\right)^{\tilde{a}_y} \left(\frac{\partial}{\partial z_A}\right)^{\tilde{a}_z} \exp(-\zeta_a r_A^2).
\end{align}
A primitive Cartesian Gaussian and a product of two primitive Cartesian Gaussians can be expressed as linear combinations of Hermite Gaussians,
\begin{align}
\label{eq:cart2herm-2}
    \phi_\mathbf{a}({\bf r})\phi_\mathbf{b}({\bf r}) = & \sum_{\tilde{p}_x=0}^{\tilde{p}_x \leq a_x+b_x} \left(E_x\right)_{a_x b_x}^{\tilde{p}_x}  \sum_{\tilde{p}_y=0}^{\tilde{p}_y \leq a_y + b_y} \left(E_y\right)_{a_y b_y}^{\tilde{p}_y} \sum_{\tilde{p}_z=0}^{\tilde{p}_z \leq a_z + b_z} \left(E_z\right)_{a_z b_z}^{\tilde{p}_z} \Lambda_\mathbf{\tilde{p}}(\mathbf{r}) \equiv \sum_\mathbf{\tilde{p}} E_{\mathbf{a} \mathbf{b}}^\mathbf{\tilde{p}} \Lambda_\mathbf{\tilde{p}}(\mathbf{r}) ,
\end{align}
with $\zeta_p \equiv \zeta_a + \zeta_b$, $\mathbf{R}_p \equiv \frac{\zeta_a \mathbf{R}_a + \zeta_b \mathbf{R}_b}{\zeta_a + \zeta_b}$,  $\zeta_q \equiv \zeta_c + \zeta_d$, $\mathbf{R}_q \equiv \frac{\zeta_c \mathbf{R}_c + \zeta_d \mathbf{R}_d}{\zeta_c + \zeta_d}$,  $E_\mathbf{a b}^\mathbf{\tilde{p}} \equiv \prod_{i={x,y,z}}\left(E_i\right)_{a_i b_i}^{\tilde{p}_i} $, and $E_\mathbf{c d}^\mathbf{\tilde{q}} \equiv \prod_{i={x,y,z}}\left(E_i\right)_{c_i d_i}^{\tilde{q}_i} $. Coefficients $E$ transforming Hermite to Cartesian Gaussians are evaluated straightforwardly by recursion\cite{VRG:mcmurchie:1978:JCP}:
\begin{align}
    \left(E_x\right)_{a_x+1 \, b_x}^{\tilde{p}_x} = & \frac{1}{2\zeta_p} \left(E_x\right)_{a_x \, b_x}^{\tilde{p}_x - 1} + (P_x - A_x) \left(E_x\right)_{a_x b_x}^{\tilde{p}_x} + (\tilde{p}_x + 1) \left(E_x\right)_{a_x b_x}^{\tilde{p}_x + 1} \\
        \left(E_x\right)_{a_x \, b_x + 1}^{\tilde{p}_x} = & \frac{1}{2\zeta_p} \left(E_x\right)_{a_x \, b_x}^{\tilde{p}_x - 1} + (P_x - B_x) \left(E_x\right)_{a_x b_x}^{\tilde{p}_x} + (\tilde{p}_x + 1) \left(E_x\right)_{a_x b_x}^{\tilde{p}_x + 1},
\end{align}
(with the obvious $y$ and $z$ analogs) and $\left(E_i\right)_{0 0 }^{0} = 1$.

The use of \cref{eq:cart2herm-2} allows to express a 4-center Coulomb integral over primitive Cartesian Gaussians as a linear combination,
\begin{align}
\label{eq:abcd_to_hermite}
[\mathbf{a} \mathbf{b}|\mathbf{c} \mathbf{d}] = & \sum_{\mathbf{\tilde{p}}, \mathbf{\tilde{q}}} E_\mathbf{a b}^\mathbf{\tilde{p}} E_\mathbf{c d}^\mathbf{\tilde{q}} [\mathbf{\tilde{p}} | \mathbf{\tilde{q}}],
\end{align}
of the Coulomb integrals between two primitive Hermite Gaussians:
\begin{align}
[\mathbf{\tilde{p}} | \mathbf{\tilde{q}}] \equiv & \,
\iint_{\mathbb{R}^3}\mathrm{d}^3\vec{r}\mathrm{d}^3\vec{r}'\,
\frac{\Lambda_\mathbf{\tilde{p}}(\vec{r}) \Lambda_\mathbf{\tilde{q}}(\vec{r}')}{\vert\vec{r} - \vec{r}'\vert} .
\end{align}
The latter can be evaluated directly,
\begin{align}
\label{eq:pq}
[\mathbf{\tilde{p}} | \mathbf{\tilde{q}}] \equiv (-1)^{l_\mathbf{\tilde{q}}} [\mathbf{\tilde{p}}+\mathbf{\tilde{q}}]^{(0)},
\end{align}
from the auxiliary integral,
\begin{align}
[\mathbf{\tilde{r}}]^{(m)} \equiv \left(\frac{\partial}{\partial x_R}\right)^{\tilde{r}_x} \left(\frac{\partial}{\partial y_R}\right)^{\tilde{r}_y} \left(\frac{\partial}{\partial z_R}\right)^{\tilde{r}_z} [\mathbf{0}]^{(m)}.
\end{align}
$[\mathbf{0}]^{(m)}$ is related to the Boys function $F_m(x)$ (or similar quantities for non-Coulomb integrals\cite{VRG:ahlrichs:2006:PCCPP}):
\begin{align}
[\mathbf{0}]^{(m)} \equiv \, & (-2 \rho)^m \frac{2 \pi^{5/2}}{\zeta_p \zeta_q \sqrt{\zeta_p+\zeta_q}} F_m(\rho |\mathbf{R}_p-\mathbf{R}_q|^2) , \\
F_m(x) \equiv \, & \int_0^1 \, \mathrm{d}y \, y^{2m} \exp(-x y^2), \label{eq:boys} \\
\rho \equiv \, & \frac{\zeta_p \zeta_q}{ \zeta_p + \zeta_q }.
\end{align}
The auxiliary integrals are evaluated recursively,
\begin{align}
\label{eq:md1}
    [\mathbf{\tilde{r}}+\mathbf{1}_i]^{(m)} = & \tilde{r}_i [\mathbf{\tilde{r}}-\mathbf{1}_i]^{(m+1)} + \left(P_i - Q_i\right) [\mathbf{\tilde{r}}]^{(m+1)},
\end{align}
starting from $[\mathbf{0}]^{(m)}$.

For high angular momenta the most expensive step in the MD4 scheme is the transformation from
Hermite to Cartesian basis, \cref{eq:abcd_to_hermite}. It is strength-reduced to a sequence of 2 transformations, for the bra and ket, respectively:
\begin{align}
\label{eq:ab_to_hermite}
[\mathbf{a} \mathbf{b}|\mathbf{\tilde{q}}] = & \sum_{\mathbf{\tilde{p}}} E_\mathbf{a b}^\mathbf{\tilde{p}} [\mathbf{\tilde{p}} | \mathbf{\tilde{q}}], \\
\label{eq:cd_to_hermite}
[\mathbf{a} \mathbf{b}|\mathbf{c} \mathbf{d}] = & \sum_{\mathbf{\tilde{q}}} E_\mathbf{c d}^\mathbf{\tilde{q}} [\mathbf{a} \mathbf{b} | \mathbf{\tilde{q}}]
\end{align}
Each transformation can be viewed and implemented as a matrix multiplication, with the bra/subscript and the ket/superscript indices denoting the matrix rows and columns (or vice versa), respectively.
Matrices $E$ and the 2-index Hermite integral matrix are dense, hence the high-performance dense matrix multiplication kernels can be used directly to implement \cref{eq:ab_to_hermite,eq:cd_to_hermite}.

Since most of the time we are interested in evaluation of integrals over (real) {\em solid harmonic} Gaussian AOs, and for $l\geq 2$ solid harmonics are less numerous than Cartesians, matrices $E$ are first contracted with the Cartesian-to-solids coefficient matrices\cite{VRG:schlegel:1995:IJQC} to produce matrices $H$ that transform from (primitive) Hermite Gaussians to (primitive) real solid Gaussians:
\begin{align}
\label{eq:}
    H_\mathbf{a b}^{\mathbf{\tilde{p}}} = \sum_{\mathbf{a}} C_{l_a m_a}^{\mathbf{a}} \sum_{\mathbf{b}} C_{l_b m_b}^{\mathbf{b}} E^{\mathbf{\tilde{p}}}_{\mathbf{a b}},
\end{align}
where $C_{l_a m_a}^{\mathbf{a}}$ is the coefficient of Cartesian Gaussian $\phi_\mathbf{a}$ in solid harmonic Gaussian $\phi_a$ with angular momentum quanta $l_a$ and $m_a$; for the detailed definitions see  Ref. \citenum{VRG:schlegel:1995:IJQC}.

\begin{center}
\begin{listing}[H]
\inputminted{c++}{md4.h}
\caption{Pseudocode for the matrix form of the MD4 approach.}
\label{listing:md4}
\end{listing}
\end{center}

The implementation of the MD4 scheme is outlined in \cref{listing:md4}.
The entire implementation involves three subkernels:
one for computing $H$ matrices [invoked once for bra $(\mathbf{ab}|$ and one for ket $|\mathbf{cd})$], one for computing the 2-index Hermite integrals $[\mathbf{\tilde{p}}|\mathbf{\tilde{q}}]$, and a matmul kernel (again invoked twice, once for the bra and once for the ket). To understand the highest angular momenta that the approach can handle let's estimate the memory footprints of all intermediate quantities involved in the evaluation. For a bra of total angular momentum $l_\mathrm{bra} = l_a + l_b$ the number of requisite Hermite Gaussians is $N_\mathbf{\tilde{p}} = (l_\mathrm{bra}+1)(l_\mathrm{bra}+2)(l_\mathrm{bra}+3)/6$ (this is known as a pyramidal number formula). For
the $[ff|ff]$ target, $N_\mathbf{\tilde{p}} = N_\mathbf{\tilde{q}} = 84$, hence in the FP64 representation the 2-index Hermite integrals occupy $84^2 \times 8 = 56,448$ bytes whereas  $N_\mathbf{\tilde{r}} = (L+1)(L+2)(L+3)/6 = 455$ (with $L=l_\mathrm{bra}+l_\mathrm{ket} = 12$), hence the 1-index Hermite integrals occupy $3,640$ bytes; for a single primitive both sets can fit into the fast memory (registers and/or scratchpad).
The matching $H$ matrices have $84 \times 7^2 = 4,116$ elements (N.B. $7$ solid harmonics in an $f$ shell) and occupy $32,928$ bytes.
However, for the $[ii|ii]$ target, the 1-index integrals already occupy 23 kB whereas the 2-index integrals and $H$ matrices occupy 1.66 MB and 615 kB, respectively. Thus for high $l$ integrals only 1-index integrals will fit into the fast memory, with the rest of data streamed in and out of the main memory.

The matrix form of MD4 is more expensive than other methods; for example, MD4 over four times
more expensive than the Rys quadrature for $[ii|ii]$ (assuming no use of horizontal recurrences in either). Nevertheless, the high FLOP count will be compensated by the efficiency of the matmul kernel on modern GPUs, even in the FP64 arithmetic.
To understand how the performance of the matrix approach will depend on the angular momentum consider a rough estimate of the arithmetic intensity (defined as the number of floating point operations performed per byte of memory transferred to the processor) of the matrix form of MD4. Since the detailed analysis is complicated we will simplify the analysis by accounting only for the FLOP count of the matrix operations (see the results in \cref{sec:performance} for justification). We will also assume that all data can be loaded into the fast memory at once. The FP64 arithmetic and primitive Gaussian AOs will be assumed throughout.
\begin{enumerate}
\item The FLOP count of the first, \begin{align}
\label{eq:make-abq}
[\mathbf{ab}|\mathbf{\tilde{q}}] = & \sum_\mathbf{\tilde{p}} H_\mathbf{ab}^\mathbf{\tilde{p}} [\mathbf{\tilde{p}}|\mathbf{\tilde{q}}],
\end{align}
and second,
\begin{align}
\label{eq:make-abcd}
[\mathbf{ab}|\mathbf{cd}] = & \sum_\mathbf{\tilde{q}} H_\mathbf{cd}^\mathbf{\tilde{q}} [ab|\mathbf{\tilde{q}}],
\end{align}
Hermite-to-Cartesian transforms
are $2 N_\mathbf{ab} N_\mathbf{\tilde{p}} N_\mathbf{\tilde{q}}$ and $2 N_\mathbf{ab} N_\mathbf{\tilde{q}} N_\mathbf{cd}$, respectively.  For contracted Gaussians the cost of evaluation of $[\mathbf{\tilde{p}}|\mathbf{\tilde{q}}]$ and the first transform grows as $K \equiv K_{bra}K_{ket}$; the cost of the second transform grows as $K_{ket} = \mathcal{O}(K^{1/2})$.
\item Hermite integrals $[\mathbf{\tilde{p}}|\mathbf{\tilde{q}}]$ are generated from the data in fast memory only and written to the main memory; the number of bytes written is $8 N_\mathbf{\tilde{p}}  N_\mathbf{\tilde{q}}$. Matrices $H$ are generated outside of the shell-quartet loops, hence their writes are not considered.
\item The number of bytes read/written in \cref{eq:make-abq} and \cref{eq:make-abcd} are $8 N_\mathbf{ab} N_\mathbf{\tilde{p}} + 8 N_\mathbf{\tilde{p}}  N_\mathbf{\tilde{q}} + 8 N_\mathbf{ab} N_\mathbf{\tilde{q}}$ and $8 N_\mathbf{ab} N_\mathbf{\tilde{q}} + 8 N_\mathbf{cd}  N_\mathbf{\tilde{q}} + 8 N_\mathbf{ab} N_\mathbf{cd}$, respectively.
\end{enumerate}
Thus we obtain the following arithmetic intensity estimate:
\begin{align}
\label{eq:AI}
  \frac{\mathrm{FLOP}}{\mathrm{byte}} = & \frac{1}{4} \times \frac{N_\mathbf{ab} N_\mathbf{\tilde{q}} \left( N_\mathbf{\tilde{p}} + N_\mathbf{cd}\right) }{2 N_\mathbf{\tilde{p}}  N_\mathbf{\tilde{q}} + N_\mathbf{ab} \left( N_\mathbf{\tilde{p}} + 2 N_\mathbf{\tilde{q}} + N_\mathbf{cd} \right) + N_\mathbf{cd} N_\mathbf{\tilde{q}} }
\end{align}
For the $[ii|ii]$ and $[ff|ff]$ targets the intensities are 16.0 and 4.2 FLOP/byte, respectively. We conclude that the performance should rise with the angular momenta of AOs. For the $[ii|ii]$ case the arithmetic intensity is {\em above} the critical value of $\sim9$ predicted for the V100 device by the Roofline model\cite{VRG:williams:2009:CA}; this {\em machine balance} value marks the transition from the memory- to compute-bound kernels on V100 when the data resides in the main memory. Thus we expect that for the $[ii|ii]$ case the matrix-based MD4 algorithm should be compute bound and can approach the hardware peak, whereas for low angular momenta the algorithm will be memory-bound. Note that the arithmetic intensity in practice will be lower than the value predicted by \cref{eq:AI} due to the limited amount of the fast memory that each kernel can exploit.
However, refining our performance model for the details of the memory hierarchy (e.g., along the lines of Ref. \citenum{VRG:yang:2020:CCPE}) is outside of the scope of this initial presentation.

\section{Implementation}

The cost of computing the Hermite-to-Gaussian transformation matrices $H$ is negligible as each bra/ket matrix is reused for many kets/bras. The reader is referred to the source code.

The performance of the 2-index Hermite integral kernel \code{compute\_pq} is limited by the memory bandwidth. To optimize the bandwidth it is necessary to maximize the occupancy, which means minimizing the fast memory footprint.  Our approach is to evaluate the 1-index integrals using \cref{eq:md1} for monotonically decreasing auxiliary indices $m$, reusing the memory occupied by $[\mathbf{\tilde{r}}]^{(m+1)}$ to store $[\mathbf{\tilde{r}}]^{(m)}$; this is akin the in-place evaluation techniques we explored in Ref. \citenum{VRG:asadchev:2023:JCTC}. The prefactors in \cref{eq:md1} and metadata (maps from index triplets $\mathbf{\tilde{r}}$ to their ordinals) are independent of $m$. Thus to maximize the reuse of this data each thread is assigned in a round-robin fashion a 1-d range of $\mathbf{\tilde{r}}$ values. Each thread then evaluates  $[\mathbf{\tilde{r}}]^{(m)}$ integrals in that range for all valid values of $m$.  Simplified pseudocode  for the 1-index integral subkernel is presented in \cref{listing:r1}.

\begin{center}
\begin{listing}[H]
\inputminted{c++}{r1.h}
\caption{Pseudocode of the subkernel evaluating $[\mathbf{\tilde{r}}]^{(m)}$ integrals.}
\label{listing:r1}
\end{listing}
\end{center}

2-index integrals  $[\mathbf{\tilde{p}}|\mathbf{\tilde{q}}]$ are then computed from 1-index integrals $[\mathbf{\tilde{r}}]^{(0)}$ via \cref{eq:pq} on the fly and written to the main memory in an order
that maximises write {\it and} subsequent read bandwidth. Each thread computes a 2-d round-robin range of 2-index integrals, one at a time, to approximately balance the load between threads. By minimizing the memory footprint of $[\mathbf{\tilde{r}}]^{(m)}$ integrals using the in-place evaluation technique of Ref. \citenum{VRG:asadchev:2023:JCTC} it is possible to evaluate 1-index integrals even for the $[ii|ii]$ integrals using only 23 kB of shared memory. This allows to assign 4 thread blocks to each SM even on the V100 GPU with relatively modest amount of  shared memory per SM and make the performance of the 2-index integral evaluation less dependent on the hardware details to ensure efficient execution on current and future generations of accelerators.  %With slight refactoring it is possible to allocate more than 48 kB to a single thread block and allow beyond the $[kk|kk]$ integrals ($L=28$, 36 kB).

Both matrix transforms can be done using batched generalized matmul (GEMM), batched over (primitive) $\mathbf{\tilde{p}}$ and $\mathbf{\tilde{q}}$ shells respectively.
The $[\mathbf{\tilde{p}}|\mathbf{\tilde{q}}]$ matrix is laid out such for a given $\mathbf{\tilde{p}}$ all  $[\mathbf{\tilde{p}}|\mathbf{\tilde{q}}]$ shell batches are contiguous in memory.  Each $[\mathbf{ab}|...]$
batch sequence is transposed to maintain batch structure for the $\mathbf{\tilde{q}} \to |\mathbf{cd})$ index transform.  This leads to $[\mathbf{\tilde{q}}|...]$ batches having strides greater than the actual dimension and worse
performance consequently.  There are a few ways to address it, for example by rearranging $[\mathbf{\tilde{q}}|\mathbf{ab})$ or
  computing Fock matrix elements directly from $[\mathbf{\tilde{q}}|\mathbf{ab})$.

Matrix multiplications are done using the NVIDIA \code{CUTLASS} library with the computational block size set small to
account for the relatively small size of individual transformations. The rest of the code relies on modern C++
metaprogramming to keep it compact and avoid the need for a separate code generator (this greatly simplifies the maintainability and usability of the library).  For example, C++ templates and compile-time-evaluated (\code{constexpr}) expressions are used
to generate index tables, Cartesian-to-solids coefficients, and so on; interested readers are referred to the source code.  The \code{compute\_pq}, the largest GPU
kernel in terms of lines of code, takes fewer than 100 lines of C++.  Porting to the AMD HIP runtime would be relatively simple as most of CUDA is abstracted through wrappers and macros.

The library interface is simple: an engine object is constructed from the given bra and ket basis sets and the integrals
are computed asynchronously for the given bra and ket index pair lists.  The lists can include shell pairs, all with the
same total contraction order (but contraction orders of individual shells may be different) and angular momenta, but each shell pair in a list must have the same total bra/ket angular momentum and contraction order.
The library does no screening, the calling code must perform the necessary screening {\em a priori}.  There are no external dependencies besides the standard CUDA and C++ toolchains.

Kernels for each target combination of total bra/ket angular momenta are automatically
generated from a single source file using the standard CMake machinery.  Each target compilation can be performed in parallel; on the benchmark machine build all kernels needed to support up to $i$ Gaussians takes a little over 3 minutes on 8 CPU cores.

\section{Performance}
\label{sec:performance}

To make the performance analysis easier and more meaningful we focus here on {\em microbenchmarking} the integral kernels, i.e., we analyze their performance for specific integral classes rather than computing e.g., the entire Fock operator matrix and/or the entire set of integrals for a given problem. While microbenchmarking is less common\cite{VRG:barca:2021:JCTC,VRG:asadchev:2023:JCTC,VRG:neese:2022:JCC} it provides a more detailed model of performance by removing extra details (e.g., screening) that can greatly influence performance of integration benchmarks. We strongly encourage others to follow suit.

The MD4 approach was implemented in the open-source \code{LibintX} library available at \textbf{\url{github.com:ValeevGroup/LibintX}} (branch \code{feature/eri/md4/v2}).
The performance of the MD4 GPU implementation was assessed against the reference \code{Libint} library\cite{VRG:valeev:2021:libint-2.7.0} compiled with \code{-march=native -mtune=native -Ofast} flags.
The \code{Libint}
kernels were executed repeatedly several hundred times with the same arguments and buffer to eliminate
the cost of initialization.

To make the comparison as favorable to the CPU code as possible,
the GPU-vs-CPU speedup is defined as the ratio of the evaluation time on a single {\em core} of the Intel Xeon Gold 6136 CPU (96 GFLOPs peak) to the evaluation time on a single PCIe-hosted V100 NVIDIA GPU (7 TFLOPs peak; note that in Ref. \cite{VRG:asadchev:2023:JCTC} we incorrectly cited the 7.8 TFLOPs figure appropriate for the SXM2-hosted V100). Evaluation of the integrals on multiple CPU cores would likely have lower efficiency due to the sharing of L2 and L3 caches by the multiple CPU cores, thermal throttling effects, etc. Thus the upper bound for the speedup for one V100 GPU vs one 12-core Intel Xeon Gold 6136 CPU can be obtained by dividing the GPU-vs-CPU-core speedups by 12.

\begin{table}
  \centering
  \begin{tabular}{c|ccc}
   \hline\hline
    Integral & $K=1$ & $K=5$ & $K=25$  \\ \hline
    $(dd|dd)$ & 93 & 58 & 45 \\
$(ff|ff)$ & 215 & 131 & 88 \\
$(gg|gg)$ & 296 & 169 & 116 \\
$(hh|hh)$ & 589 & 293 & 172 \\
$(ii|ii)$ & 1054 & 415 & 182
 \\
    \hline
  \end{tabular}
  \caption{Relative speedup (see text for details) of the \code{LibintX} implementation of the MD method executing on 1 NVIDIA V100 GPU vs. the reference \code{Libint} implementation of the OSHGP method running on one core of Intel Xeon Gold 6136 CPU. The GPU-to-CPU ratio of hardware peak FP64 FLOP rates is 73, hence the values greater than 73 indicate when the GPU implementation is more efficient than the CPU counterpart. $K$ denotes the total contraction degree of the target integrals.}
  \label{tab:abcd-perf}
\end{table}

\Cref{tab:abcd-perf} presents the results of comparison.
The ratio of peak FLOP rate of the V100 GPU to that of the 1 CPU core is
73:1, which can be used to as the baseline for judging the relative efficiency of the GPU code vs the reference CPU code.
The uncontracted case ($K=1$) exhibited the highest GPU/CPU speedup. This is not surprising since the CPU Obara-Saika-based Head-Gordon-Pople (OSHGP) implementation in \code{Libint} leverages the horizontal recurrence relation to greatly reduce the cost for high contraction degrees by shifting much work outside of the contraction loops. Indeed, with the increasing contraction degree the speedups decrease due to the cost of HRR being amortized over several VRR steps. The performance decreases with
angular momentum due to the increasing relative cost of the 2-index Hermite integrals and the increasing algorithmic
efficiency of the Obara-Saika scheme. The greatest speedup, over 1000 times, is achieved for the uncontracted
$(ii|ii)$ integral; this greatly exceeds the speedup target of 73. Only for the contracted $(dd|dd)$ integrals the GPU/CPU performance ratio failed to exceed the speedup target.
In our future work we will introduce refinement of the algorithm targeted at the lower angular momentum integrals [$(ff|ff)$ and below].

\begin{table}
  \centering
  \begin{tabular}{c|cc|cc}
   \hline\hline
    & \multicolumn{2}{c|}{$(dd|dd)$} & \multicolumn{2}{c}{$(ii|ii)$}  \\ \hline
    & FLOPs \% & Cycles & FLOPs \% & Cycles \\ \hline
    $H_{ab}^\mathbf{\tilde{p}}$ & 2 &  $<1$ & 5 &  $<1$ \\
$H_{cd}^\mathbf{\tilde{q}}$ & 2 &  $<1$ & 5 &  $<1$ \\
$[\mathbf{\tilde{p}}|\mathbf{\tilde{q}}]$    & 6 &  18 & 4 &  11 \\
$(ab|\mathbf{\tilde{q}}]$   & 69 & 37 & 75 & 48 \\
$(ab|cd)$  & 23 & 43 & 32 & 40
 \\
    \hline
  \end{tabular}
  \caption{Performance breakdown of the MD4 implementation by subkernel for two uncontracted ($K=1$) integral types. For each subkernel invocation the fraction of the peak FLOP rate and the percentage of the total clock cycles are listed.}
  \label{tab:perf}
\end{table}

\Cref{tab:perf} presents a closer look at the performance and costs of the MD4 implementation for the uncontracted $(dd|dd)$ and $(ii|ii)$ integrals.
As expected, the cost of computing the $H$ matrices accounts for less than 1\% of the total execution time.
Evaluation of the Hermite basis integrals is almost entirely memory-bounded but accounts for the decreasing fraction of the total cost as the
angular momentum increases. Most of the time ($ > 80\%$) is spent in the two matrix transformations, with the second
transformation being less efficient. The overall FLOP rate for the uncontracted $(dd|dd)$ and $(ii|ii)$ integrals are 36\% and 49\% of the peak 7 TFLOPs, respectively. This is the highest reported figure for high angular momentum Gaussian AO integrals we have seen to date.

For contracted integrals ($K>1$) the distribution of FLOPs among the subkernels will change slightly, since the FLOP counts of  $[\mathbf{\tilde{p}}|\mathbf{\tilde{q}}]$ and $(\mathbf{ab}|\mathbf{\tilde{q}}]$ kernels are proportional to $K$, with the rest scaling as $K^{1/2}$. What does not change, however, is the FLOP dominance of the GEMM-based subkernels. Thus we expect similar absolute FLOP rate to be achieved for both contracted and uncontracted integrals.

\section{Conclusion}
Despite being more expensive, the matrix multiplication form of the McMurchie-Davidson scheme executed on an NVIDIA V100 GPU achieves significant, up to 3
orders of magnitude, speedups over the reference Obara-Saika-based implementation on a CPU core; this greatly exceeds the 73:1 ratio of the peak FLOP rates of the GPU to the CPU. This work illustrates how to remove the barriers to
computing high-angular momentum two-electron four-center integrals on GPU, opening doors for further
development of computational chemistry on GPUs and other accelerators.

The approach presented here will be refined
and adapted to better handle lower angular momentum, high contraction integrals as well as the
three-center integrals. Evaluation of the exchange matrix using our integral engine will be also presented elsewhere. The code is distributed as a part of open-source C++ {\tt LibintX} library under
LGPL3; it is available publicly at \url{github.com:ValeevGroup/LibintX}.

\begin{acknowledgement}
We would like to congratulate Prof. Lindh on his numerous major contributions to our field. In particular, his work on the integral evaluation algorithms\cite{VRG:lindh:1991:JCP,VRG:ryu:1991:CPL,VRG:reine:2011:WIRCMS}, on parallel algorithms for quantum chemistry\cite{VRG:rendell:1992:CPL}, and on integral tensor approximations\cite{VRG:aquilante:2008:JCP,VRG:pedersen:2009:TCA,VRG:merlot:2013:JCC} has strong overlap with our work here and elsewhere and continues to be as relevant as ever. Congratulations, Roland!

AA would like to thank the author of Ref. \citenum{VRG:samu:2018:} for creating such a useful reference document.

This work was supported by the U.S. Department of Energy award DE-SC0022263 provided via the Scientific Discovery through Advanced Computing (SciDAC) program by the Offices of Advanced Scientific Computing Research (ASCR) and Basic Energy Sciences (BES). We also acknowledge Advanced Research Computing at Virginia Tech (www.arc.vt.edu) for providing computational resources and technical support that have contributed to the results reported within this paper.
\end{acknowledgement}

\bibliography{vrgrefs}

%\newpage
%\thispagestyle{empty}
%\includegraphics[width=\textwidth]{toc.tiff}

\end{document}